\documentclass[pdflatex,sn-mathphys]{sn-jnl}

\jyear{2021}%

\theoremstyle{thmstyleone}%
\theoremstyle{thmstyletwo}%

\theoremstyle{thmstylethree}%

\begin{document}

\title[Frequency chirped Fourier-Transform spectroscopy]{Frequency chirped Fourier-Transform spectroscopy}

\author*[1]{\fnm{Sergej} \sur{Markmann}}\email{msergej@phys.ethz.ch}

\author[1]{\fnm{Martin} \sur{Francki{\'e}}}

\author[1]{\fnm{Mathieu} \sur{Bertrand}}

\author[1]{\fnm{Mehran} \sur{Shahmohammadi}}

\author[1]{\fnm{Andres} \sur{Forrer}}

\author[1]{\fnm{Pierre} \sur{Jouy}}

\author[1]{\fnm{Mattias} \sur{Beck}}

\author[1]{\fnm{J{\'e}r{\^o}me} \sur{Faist}}

\author[1]{\fnm{Giacomo} \sur{Scalari}}

\affil[1]{\orgdiv{Institute for Quantum Electronics}, \orgname{ETH Z{\"u}rich}, \orgaddress{\street{Auguste-Piccard-Hof 1}, \city{Z{\"u}rich}, \postcode{8093}, \state{Z{\"u}rich}, \country{Switzerland}}}

\abstract{Fast 
(sub-second) spectroscopy with high spectral resolution is of vital importance for revealing quantum chemistry kinetics of complex chemical and biological reactions. Fourier transform (FT) spectrometers can achieve high spectral resolution and operate at hundreds of ms time scales in rapid-scan mode. However, the linear translation of a scanning mirror imposes stringent time-resolution limitations to these systems, which makes simultaneous high spectral and temporal resolution impossible. Here, we demonstrate an FT spectrometer whose operational principle is based on continuous rotational, rather than linear, motion of the scanning mirror, decoupling the spectral resolution from the temporal one. This enables 0.5 cm${}^{-1}$ resolution on sub-ms time scales. Furthermore, we show that such rotational FT spectrometers can perform dual-comb spectroscopy with a single comb source, since the Doppler-shifted version of the comb serves as the second comb.
In this way, we combine the advantages of dual-comb and FT spectroscopy using a single quantum cascade laser frequency comb as a light source. Our technique does not require any diffractive or dispersive optical elements and hence preserve the Jacquinot's-,  Fellgett's-, and Connes'-advantages of FT spectrometers. The system supports a large optical bandwidth from visible to THz frequencies. The combination of a rotational delay line with collimated coherent or non-coherent light sources pave the way for FT spectrometers in applications where high speed, large optical bandwidth, and high spectral resolution are desired.}


\maketitle

\section{Introduction}\label{sec1}
Real-time monitoring of non-repetitive events is currently a rapidly developing area in the field of instrumentation \cite{schmidt_chapter_2020}. Rapid optical analysis in applications such as chemical reaction monitoring \cite{fleisher_mid-infrared_2014}, protein dynamics \cite{norahan_microsecond-resolved_2021}, micromachining \cite{cheng_review_2013} or phase transitions in novel material systems \cite{ratri_phase-transition_2013} is of uttermost importance to understand the underlying complex dynamics. A general purpose instrument for spectroscopic characterisation is a Fourier-Transform (FT) spectrometer. However, the scan rate of the traditional FT spectrometer is often too slow when compared to the timescale of the physical processes of interest \cite{wagner_time-resolved_2011}. This is due to the fact that the information is acquired via the linear displacement of a mirror, which results in a trade-off between the temporal and spectral resolutions given that the system is not limited by the signal-to-noise ratio. In the last two decades two different directions appeared in the technological development of spectrometers. On one hand, the miniaturisation of spectrometers \cite{yang_miniaturization_2021} which is very promising for cost-effective lab-on chip systems \cite{souza_fourier_2018,zheng_microring_2019,yang_miniaturization_2021,fathy_-chip_2020,li_-chip_2021,zhang_parametric_2013} for specialized industrial applications. On the other hand, several new operational principles have been proposed and demonstrated with the goal of eliminating the above-described disadvantages of traditional benchtop FT spectrometers. Recently, successful proof-of-concept of phase-controlled FT spectroscopy \cite{hashimoto_phase-controlled_2018-1} and time-strech spectroscopy \cite{kawai_time-stretch_2020,goda_dispersive_2013,mahjoubfar_time_2017} have been demonstrated, which outperform the acquisition speed of traditional FT spectrometer by several orders of magnitude, while maintaining a relatively high spectral resolution. However, these newly developed methods rely on dispersive elements such as gratings, dispersive fibers, or prisms. Hence, they can be operated only over a limited optical bandwidth, mainly at visible and near-infrared frequencies. Although these spectral regions are of interest for certain applications, the most relevant one for molecular spectroscopy is the so-called molecular fingerprint region which covers mid-infrared (mid-IR) wavelengths from 4 to 12 $\mu$m. Recent progress in dual-comb spectrometers with mid-IR quantum cascade lasers (QCLs) \cite{hugi_mid-infrared_2012} have gained a lot of attraction. Such a system acquires information not only on microsecond times scales, but can also be used for high-resolution spectroscopy via spectral interleaving \cite{villares_dual-comb_2014} with high brightness which is very important for highly absorbing samples. However, there is a fundamental issue, which is the requirement on coherence for both frequency combs. The lack of mutual coherence of two free running lasers results in additive noise and the absence of an absolute frequency reference \cite{coddington_dual-comb_2016}. The latter can be overcome, as has been recently shown in \cite{gianella_frequency_2022}, at the cost of additional spectrometer complexity. These technological hurdles raise the question of whether it is possible to perform dual-comb spectroscopy with a single frequency comb source, without sacrificing spectral resolution and still keeping an acquisition speed sufficient for the majority of high-speed applications (typically 0.1-100's of ms).\\
This question naturally leads one to consider the incorporation of the frequency comb source into a FT spectrometer with the goal to preserve the advantages of a dual-comb system and the flexibility of the FT spectrometer. Such integration has been realized in ref. \cite{mandon_fourier_2009}. Even though the comb modes (mode spacing of 140 MHz) were not resolved in this work due to the limitation of the mechanical delay line (resolution of 1.5 GHz), and the interferogram acquisition was only slightly below 5 minutes, it demonstrated the high potential of this approach. \\
To address the growing demand for real-time spectroscopy in the NIR and mid-IR (2um-14um), the temporal and spectral performance of existing FT spectrometers has to be improved. Several optical delay line systems have been developed which are based on a rotational delay line, such as a circular involute stage \cite{xu_circular_2004} and curvilinear reflectors \cite{skorobogatiy_linear_2014}, \cite{kim_high_2008}. These two systems have in common that the optical path difference is angle dependent, obtained via rotational optics which guide the incoming beam to a static reflector which self reflects the beam back. Hence, in such systems the incoming and out going beam are anti-parallel, which will result in large undesired feedback into the laser source. Moreover, these systems do not support extended light beams and hence posse a phase-dependent-delay across the beam diameter, which make them impractical for usage in the FT spectrometer. On the other hand, a compact flexible multi-pass rotary delay line using spinning micro-machined mirrors have been developed \cite{fleddermann_compact_2017}. The big disadvantage of these systems is relatively small achievable optical path difference, which is typically less than 1 mm, and the requirement of fiber optics and optical circulators, thus limiting this system to a narrow optical bandwidth. Another interesting approach was shown in ref. \cite{kim_novel_2010} where a rotating retro-reflector in combination with a static plain mirror is used. The incoming beam which is guided via a rotational retro-reflector to the plane mirror is back reflected. The disadvantage of this system lies in a not separable incoming and outgoing beams. Moreover, the system size grows linearly with the increase of the optical achievable delay path. This make the system impractical for usage of desired delay lengths of 3 cm and more at large acquisition speeds.\newline
By taking the advantages and disadvantages of the above described optical delay systems into account, we can define criteria which the desired system should fulfill: (I) Support of a large optical bandwidth; hence dispersive elements have to be avoided. (II) High spectral resolution; 0.5 cm$^{-1}$ or lower. (III) Preserved polarisation of the light source. (IV) Fast acquisition speeds: 0.1-100's of ms. (V) Easy integration into existing systems and hence small footprint. (VI) Decoupled temporal and spectral resolution as well the decoupling of the incoming and outgoing beam. (VII) Support of extended beam diameter size.\\ 
In this work we demonstrate an FT spectrometer based on a rotational delay line and a single frequency comb light source, which fulfills the conditions listed above and overcomes many of the difficulties of dual-comb and linear FT spectroscopy. In conventional FT spectrometers, which are based on linear or pendulum motion of a scanning mirror, the direction of the moving mirror has to be reversed. This sets the maximum achievable acquisition speed for a given spectral resolution, due to the mirror inertia which would require large external forces that scale quadratically with the acquisition time. The use of a rotational element decouples the spectral resolution from the interferogram acquisition time. This is due to the fact, that the spectral resolution is fixed and is given by the total optical path length of the rotational delay line. The temporal resolution is given by the angular velocity of the rotational optical element and is independent on the total optical path length. With our rotational FT spectrometers we can reach ms acquisition speeds which can be improved by further technological development as will be discussed later. Such a high acquisition speed results in a high signal-to-noise-ratio (SNR), due to minimization of 1/f noise within the signal bandwidth which is crucial for a free running mid-IR QCL \cite{cappelli_intrinsic_2015}, \cite{tombez_linewidth_2012}. The combination of our rotational FT spectrometer with the frequency comb as a light source does not limit our system in spectral resolution given by the total optical path length. As a proof-of-principle, by exploiting a quantum cascade laser frequency comb as a light source, we demonstrate high spectral resolution measurements via interleaving \cite{gianella_high-resolution_2020}, resolving the  transmission resonances of a silicon etalon and the absorption lines of a Doppler-broadened low-pressure methane sample (CH$_4$) with a resolution $<$ 250 MHz. Moreover, we demonstrate combined high spectral and temporal spectrometer capabilities by monitoring the absorbance lines of a low pressure methane during evacuation of a gas cell.

\section{Results}\label{sec2}
\subsection{Rotational FT-spectrometer}\label{subsec2}
\begin{figure}[h]
    \centering
    \includegraphics[width=4.5in]{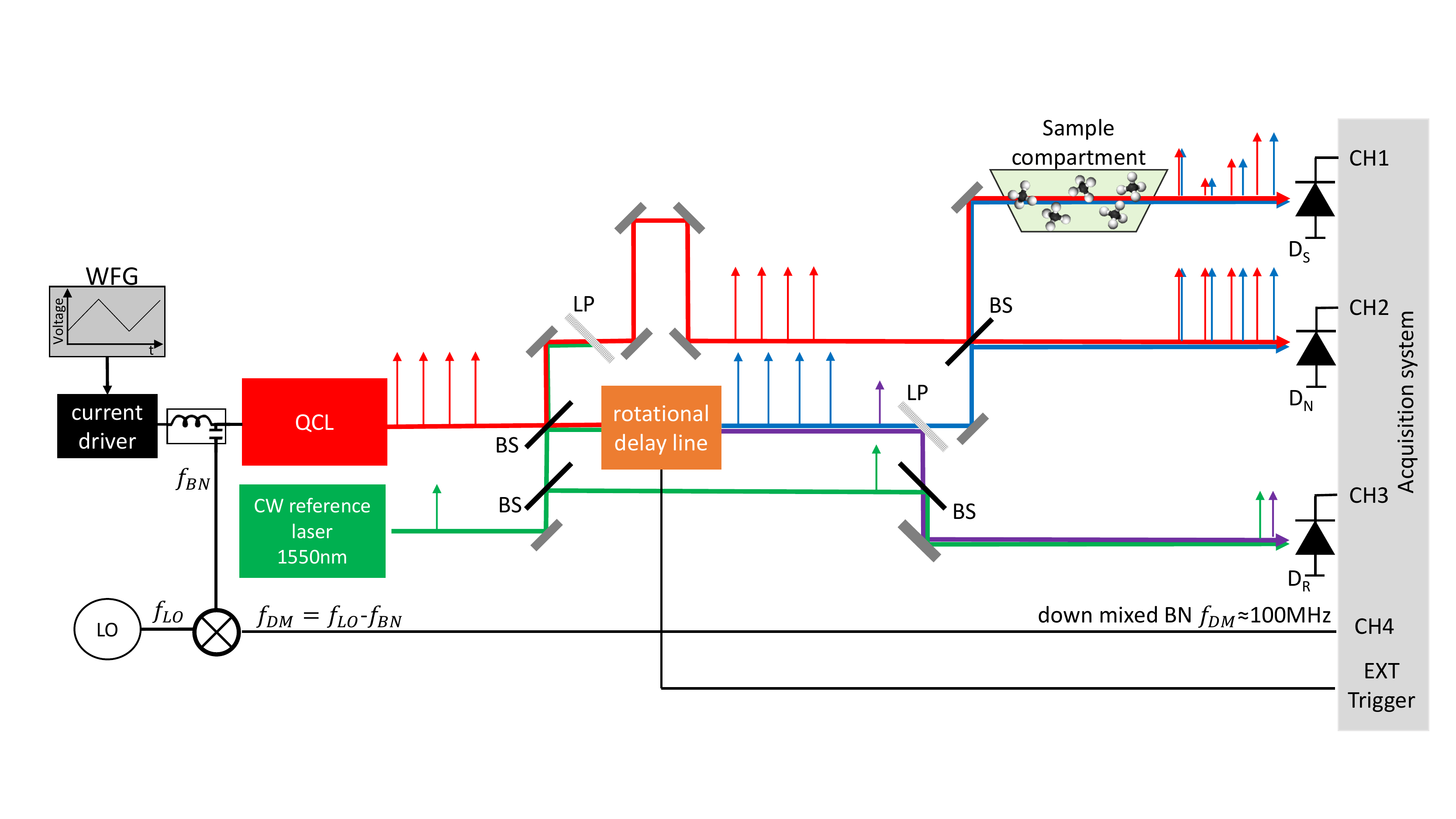}
    \caption{\textbf{Schematic of the constructed rotational FT spectrometer with a QCL as a light source. A CW reference laser which is used to provide a frequency reference and serves also as a tool for mapping out and removing the induced frequency chirp imposed by the nonlinear rotational delay line. QCL and CW laser beam are spatially superimposed via a beam splitter (BS) and also separated via the optical low-pass filter (LP). Interferograms of the QCL are acquired with the sample (D$_S$) and normalization (D$_N$) detector, where the reference interferogram of the CW laser is recorded with detector D$_R$. All measurements are synchronized with the trigger from the rotational delay line. The wave function generator (FWG) can be used for continuously tuning of the QCL operation point by applying a ramp to the current driver of the QCL. The inter-modal beat note signal can be extracted directly from the waveguide via bias-tee and the down-mixed version of it $f_{DM}$ can be recorded with the help of the local oscillator LO.}}
    \label{fig:experiment}
\end{figure}

\begin{figure}[h]
    \centering
    \includegraphics[width=5.2in]{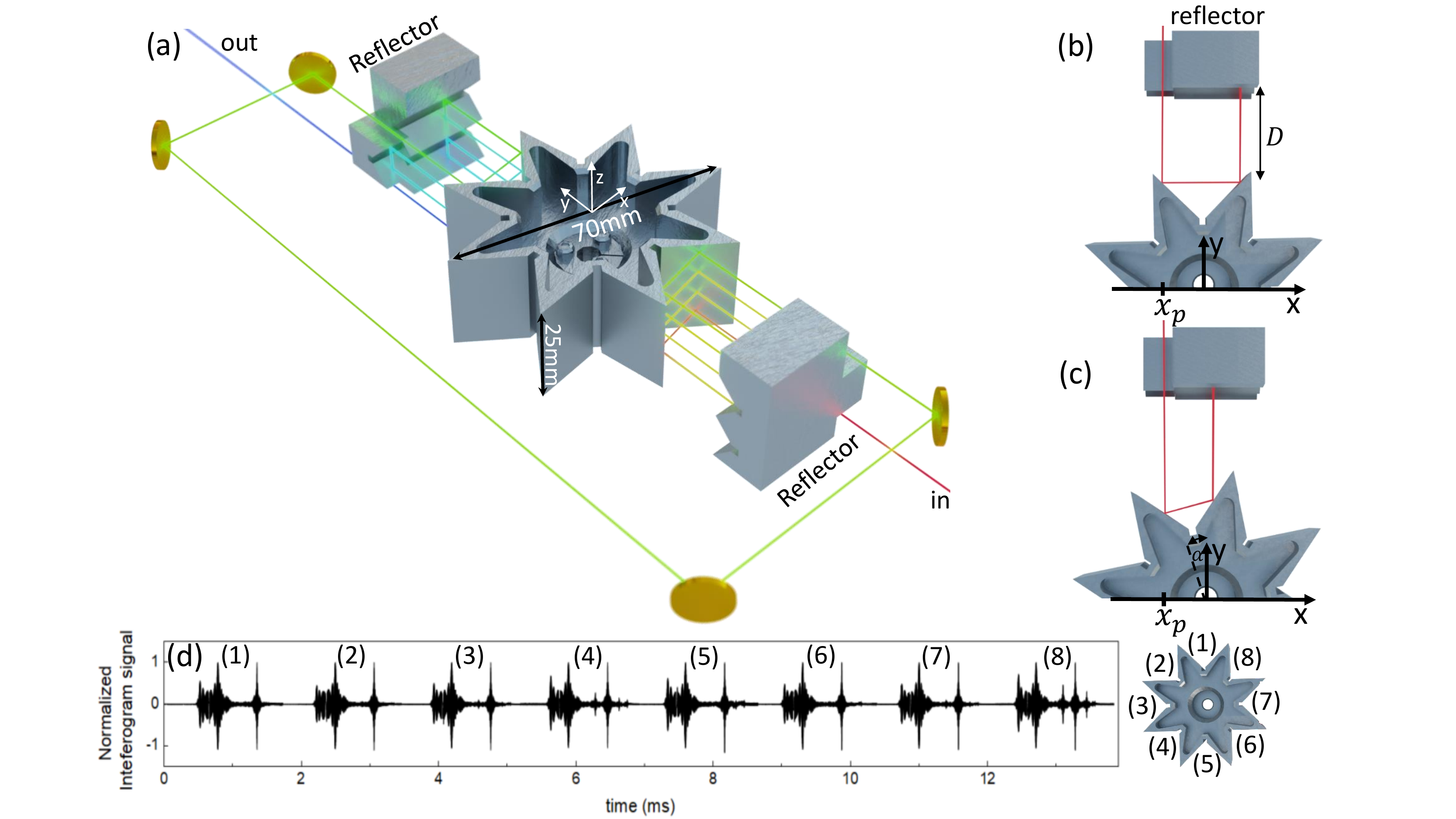}
    \caption{\textbf{Rotational delay line. (a) Realization of the rotational delay line (RDL) consisting out of three-dimensional rotational retro-reflecting (RR) octagrammic prism and a static retro-reflecting system (SR). The SR is arrange in such a way with respect to RR, that multiple reflections between SR and RR are possible. By exploiting the point symmetry of the system, the optical path delay can be doubled with two SR on opposite sides of the RR.    
    (b)/(c) Top view of a beam propagation in the RDL at rotation angle 0$^\circ$ and $\alpha$. (d) Recorded interferogram on each rotational retro-reflector of delay line.}}
    \label{OPD1}
\end{figure}

A diagram of the rotational-based FT spectrometer is shown in Fig.~\ref{fig:experiment}. In this first demonstration we employ a mid-IR QCL frequency comb \cite{hugi_mid-infrared_2012} as a light source emitting around 8 $\mu$m wavelength. However any collimated light source with beam diameter below 7 mm could be used in our spectrometer. The QCL is 2.5 mm long (free spectral range of $\approx$ 19.7 GHz) with high-reflection coating on the back facet and with maximum emission power of 60 mW at T=-20$^{\circ}$C. In contrast to amplitude modulated frequency comb light sources, which emit light pulses spaced by the cavity round-trip time ($t_{rep}$=$\frac{1}{f_{rep}}$), the QCL emits a frequency modulated comb (FM-comb). This type of a frequency comb exhibits a linear phase difference between neighboring longitudinal laser modes, resulting in a linearly chirped instantaneous frequency with nearly constant intensity \cite{singleton_evidence_2018}. The absolute frequencies $f_m$ of the comb teeth in the optical domain are described by $f_m=f_{ceo}+mf_{rep}$, where $f_{ceo}$ is the carrier envelope offset frequency, $f_{rep}$ the repetition rate and $m$ an integer representing the mode number. The emitted and collimated QCL light beam is divided by a beam splitter (70/30) into two paths; one static beam path and one which passes through the rotational delay line (RDL). The two beams are then recombined on a second beam splitter (50/50), creating a sample and a normalization beam path. Since the beam is going through the RDL, it acquires an optical path delay (OPD) with respect to the static beam, two separate interferograms can be recorded at the sample ($D_S$) and normalization ($D_N$) detectors, respectively. We use thermo-electrically cooled HgCdTe detectors from Vigo system (model:PVMI-4TE-10.6-1x1-MIP-10k-100M-TO8-wZnSeAR) with a peak sensitivity at 8 $\mu$m and 100 MHz bandwidth. As for dual-comb spectroscopy setups, the normalization detector is required for canceling common mode fluctuations which can originate from power and frequency instabilities of the QCL comb. The key element of our setup is the RDL which is schematically shown in Fig.~\ref{OPD1} (a). It consists of a rotational retro-reflecting (RR) octagrammic prism and a static retro-reflecting (SR) system, which are arranged so that the incoming beam towards the SR is redirected along the rotation axis (z-axis) and reflected back towards the RR. The RR can in principle consist of any number $M\geq 2$ retro-reflectors, symmetrically arranged around the rotational axis z. The optimal $M$ will depend on the repetition rate of the laser, the speed of the motor, the tolerable losses due to number of reflection in the delay line, and the desired acquisition rate. In our case, $M = 8$ is optimal for the target of recording a single IFG in $\sim$ 1 ms and faster. The SR consists of two retro reflectors arranged next to each other and shifted along the z-axis, so that the incoming beam is reflected several times at different height of the RR. In this configuration, the incoming and out-going beams on the same side of the RR have opposite propagation directions and are separated in height. Moreover, the retro-reflecting arrangement preserves the polarisation of the laser beam. The out-going beam can be guided via mirrors to the opposing side of the RR in order to bring it to the height of the initial incoming beam. This way of lowering the beam has the additional advantage of doubling the optical path difference, which allows for resolving the comb modes. Fig.~\ref{OPD1} (b)-(c) illustrate the beam propagation at zero degree and at an angle $\alpha$. The general expression for the total optical path length in the RDL is given by
\begin{equation} 
\begin{split}
L(\alpha)=& 2N(D+S_1x_p-C_1+\sqrt{(T-x_p)^2+(S_3T+C_3-S_1x_p-C_1)^2}\\
          & -S_3T-C_3)+2(N-1)l_4,
\label{eqn:path}
\end{split}
\end{equation}
where $S_1=\tan(\alpha-\pi/4)$,
 $C_1=\cos(\alpha)b+S_1\sin(\alpha)b$,
 $S_2=\tan(\alpha+\pi/4)$,
 $C_2=\cos(\alpha)b+S_2\sin(\alpha)b$,
 $S_3=\tan(2\alpha)$ and 
 $C_3=C_1+S_1x_p-S_3x_p$.
Here, $\alpha(t)=\omega t$ is the rotation angle, $\omega$ the angular velocity of the RR, $t$ is the laboratory time, and $N$ is the number of times the beam is injected onto the RR (in our case $N=4$). The derivation of Eq.~\eqref{eqn:path} can be found in the Supplementary Materials.
The OPD is then $\Delta L=L(\alpha)-L(\alpha=0)$. Fig.~\ref{OPD2} (a) shows the optical path length according to Eq.~\eqref{eqn:path} when the incoming beam enters the RDL at $x=x_p$ and at $x=x_p+\frac{d}{2}$ at $\alpha(t=0)=0$), where $d$ is the beam diameter. It shows that the incoming position of the beam on the $x$-axis (Fig. ~\ref{OPD1} (b) and (c)) determines the maximum achievable OPD. It is important to note, that the phase delay is independent on the laser beam diameter $d$, which is crucial for spectroscopy applications. The beam diameter size only influences a maximum achievable OPD. The maximum achievable OPD is given for $d=0$ and the entry point $x=x_p$. Moreover, the phase delay is nonlinear in the angle $\alpha$, and hence in the laboratory time t.\\
In order to take the nonlinear OPD into account, as well as to provide an absolute frequency reference for the QCL spectrum, a single mode continuous-wave (CW) reference laser (Rio GRANDE 1550 nm High Power Laser Module) is spatially superimposed on the QCL beam. In this way, the two beams are co-propagating through the RDL and experience the same nonlinear OPD. After propagation through the RDL the reference beam is spatially separated from the QCL one by means of an optical low-pass filter, which acts as a reflective element for the reference laser and transmissive for mid-IR QCL beam. As in the case of the QCL, a static and a non-static CW reference laser beam are combined on a beam splitter and the resulting interferogram is recorded on the reference detector $D_r$.\\ 
In addition to the detection of the optical signals, we exploit the fact that the beating of the linearly spaced frequency comb modes of the QCL can be directly measured in the radio-frequency domain. The beating gives rise to a modulation of the laser bias and can be extracted from the device through a bias-tee (Fig.\ref{fig:experiment}). We record this inter-modal beat note signal by down-mixing (DM) it with a local oscillator LO with the frequency $f_{LO}$ to the frequency $f_{DM}\approx$ 100 MHz. All signals are then digitized with the acquisition card (model: Spectrum M4i.445x-x4) with 500 MS/s and 14 bit resolution. The digitized data is then band pass filtered with the goal to suppress the noise outside the signal bandwidth. With the help of the LO and the recorded $f_{DM}$ signal, the repetition rate $f_{rep}= \lvert f_{LO}-f_{DM} \rvert$ of the QCL is determined. The optical frequency of the reference CW laser was determined with precision better than few 100s of MHz using an optical spectrum analyser. This reference laser features frequency stability of $<$ 15 kHz. Since the zero crossings of the CW reference interferogram are used for resampling of the QCL interferogram, the frequency accuracy of the resampled QCL spectrum is limited by the optical frequency resolution of the CW reference laser which is several order of magnitudes larger than the laser stability of $<$ 15 kHz. However, the frequency of the reference laser can be calibrated more precisely (several MHz) with a narrow absorption line of a known reference gas, as also realized in this work in the interleaving section. Please note, that a laser with a frequency stabilty of 1-2 MHz would be also sufficient for our experiment since we perform a calibration on narrow absorption line. The developed RDL provide around 3 cm of OPD. A maximal RR angular frequency of 85 Hz (limited by the employed motor's stability) results in relatively high acquisition rates, with a duty cycle of 50$\%$ as shown in Fig.~\ref{OPD1} (d), where each interferogram is recorded on one of the retro-reflectors of the octogram. Moreover, as we will discuss later any imperfections in the fabrication of the RR might result in tiny differences in the interferogram acquired on each retro (as can be seen in Fig.~\ref{OPD1} (d) (1) and (8)). However, since the normalization detector is used, the important measure is the spectral ratio of the sample and the normalization detector which will eliminate any fabrication imperfections of the RR.      
\subsection{Operational principle of rotational FT spectrometers}
\begin{figure}[h]
    \centering
    \includegraphics[width=1\textwidth]{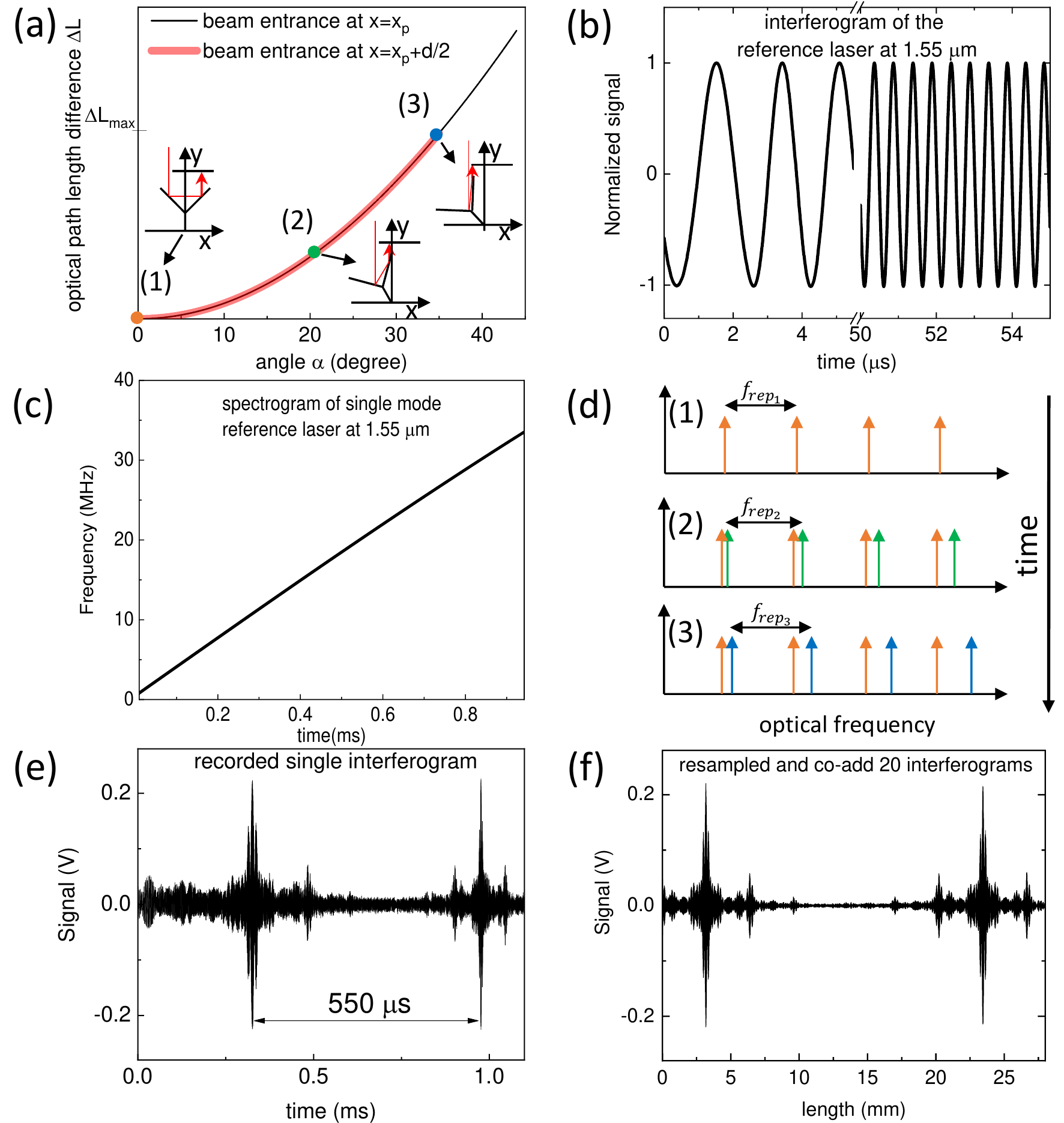}
    \caption{\textbf{Operational principal. (a) Induced nonlinear optical path length difference $\Delta$L by rotational delay line as a function of rotation angle $\alpha$. The maximal achievable path length difference depends on the entry point of the light beam along the x-axis (Fig.~\ref{OPD1}(b)/(c)). The orientation of the rotational delay line with respect to the incoming beam is shown for the selected point. (b) Interferogram time slices of CW reference laser on detector D$_R$. (c) Spectrogram of CW reference laser which shows almost linear induced frequency chirp by rotational delay line. (d) Illustration of induced frequency chirp to a frequency comb. Three different color coded time points of comb spectrum, which corresponds to the same color-coded points on the curve in (a).(e) Recorded QCL interferogram on one of eight retro-reflectors of the octogram. Increasing frequency chirp is clearly visible with increasing time. (f) Resampled QCL inteferogram of (e) on zero crossings of a CW reference laser from (b) with a  coherently co-added 20 consecutive interferograms.}}
    \label{OPD2}
\end{figure}

Since the individual interferogram is acquired within a ms or even faster, this has implications on the signal itself due to the Doppler effect \cite{benirschke_frequency_2021}. The relativistic Doppler effect shifts the frequency $f_m$ of an electromagnetic wave to the frequency $f_{D_m}$ according to 
\begin{equation} 
f_{D_m}=f_m\sqrt{\frac{1-v/c}{1+v/c}} \approx f_m(1-v/c),
\end{equation}
where $v = \frac{dL_{tot}}{dt} << c$ is the rate of change of the optical path length $L_{tot}$ and $c$ is the speed of light. The frequency shift depends on the speed of the motor and the OPD which is given by the radius of the rotating disc, and the number of reflections onto the rotating disc (see Supplementary materials), which in our case gives an average $v$ of about 40 m/s. The Doppler effect of mid-IR frequencies at these speeds is significant and amounts to several MHz. Since the Doppler-shifted and non-Doppler shifted light beams are superimposed onto a detector, the neighboring optical frequencies of the beams will beat and generate a heterodyne signal in the radio frequency (RF) domain given by
\begin{equation} 
f_{RF_m}(t)= f_m \frac{1}{c}\frac{\text{d}L_\text{tot}(\alpha(t))}{\text{d}t}  = f_m \frac{\partial L_\text{tot}(\alpha)}{\partial \alpha}\frac{\omega}{c}.
\label{eq:fRFi}
\end{equation}
The frequency chirp induced by the nonlinear OPD \ref{OPD2} (a) is visualised in Fig. \ref{OPD2} (b) by recording an interferogram of the single-mode reference laser. Here, the first 5 $\mu$s of the recorded interferogram and the same time interval after 50 $\mu$s is shown, while the chirped Doppler shift of the entire interferogram is shown by the spectrogram in Fig. \ref{OPD2} (c). 

In the case of a frequency comb source, the signal recorded on detector $D_S$ and $D_N$ without a sample is given by 
\begin{equation} 
S=\Re \left(\sum_{m=0}^{\infty}A_m\exp\left(\text{j}\left( 2\pi f_{RF_m}(t) t+\phi_m\right)\right)\right)
\label{eq:SRFi}
\end{equation}
where $A_m$ is the amplitude and $\phi_m$ the phase of mode $m$. The nonlinear OPD will not only induce a chirp to all modes $m$, resulting in absolute frequency shifts of the mode frequencies $f_{RF_m}$, but will also directly modify $f_{rep}$ of the comb during the interferogram acquisition. This effect is illustrated in Fig.~\ref{OPD2} (d) at three different time points, corresponding to the same color-coded points on the curve in Fig.~\ref{OPD2} (a). The first orange point at $\alpha(t=0)=0$ is related to Fig.~\ref{OPD2} (d) (1) and corresponds to the case with zero OPD and no Doppler shift. Increasing $\alpha$ to 20 degrees impacts the repetition rate of the comb ($f_{rep_2}=f_{rep_1}\cdot(1+v(\alpha=20^\circ)/c)$) as well as the absolute frequencies of the individual optical modes, which acquire a Doppler shift as well (Fig.~\ref{OPD2} (d) (2)). A further angle increase of the RR results in a larger Doppler shift and hence also larger $f_{rep_3}$ (Fig.~\ref{OPD2} (d) (3)). Such a frequency-chirped interferogram of a QCL frequency comb is shown in Fig.~\ref{OPD2} (e). In order to obtain a chirp-free interferogram from which the spectrum can be retrieved, the chirped QCL interferogram is resampled using the zero crossings of the reference laser interferogram. This is possible since both beams undergo identical nonlinear delay and the frequency response of the electronic is flat. The resampled interferogram of the QCL frequency comb with improved SNR by coherently averaging 20 interferograms, which averages out uncorrelated noise present in individual interferograms is shown in Fig.~\ref{OPD2} (f). The resolution of the system is 15 GHz and is given by the maximum OPD of 3 cm, while the single burst acquisition speed on one of the octogram retro-reflectors is 1 ms which is given by the maximum angular velocity of the motor and the geometrical realisation of the rotating and static retro-reflectors via Eq.~\eqref{eqn:path}.

\subsection{Spectroscopy with rotational FT spectrometer}
In order to perform absorption spectroscopy on a sample, a sequence of two measurements is needed: a background spectrum and sample spectrum. The background spectrum is acquired on the sample $D_S$ and normalization detector $D_N$ without a sample present. Since the QCL spectrum is extracted from the acquired single-burst interferogram with the knowledge of the computed $f_{rep}$, we employ sub-nominal resolution method described in ref. \cite{rutkowski_optical_2018}, in which precisely one repetition rate of the laser is acquired, in order to eliminate spectral leakage. As pointed out earlier, the normalization detector $D_N$ is used to suppress common noise on the two beam paths, stemming from laser relative intensity noise (RIN). Please note, that the laser RIN of a free running QCL significantly reduces for acquisition speed of 1ms for an interferogram \cite{tombez_linewidth_2012}. This is also where the speed of our rotational FT spectrometer plays an important role. Moreover, we have also verified that the light scattering due to surface quality (diamond polish aluminum with surface roughness $<$10 nm Ra, arithmetical mean deviation) of the RDL (RR and RS) does not contribute to the intensity noise. This is realized by computing the background ratio 
\begin{equation}
    R_{BG}=\frac{S^{BG}_{S}}{S^{BG}_{{N}}},
\end{equation}
where $S^{BG}_{{S}}$ and $S^{BG}_{{N}}$ are the spectra recorded by detector $D_S$ and $D_N$, respectively. Placing a sample into the sample beam path (Fig. \ref{fig:experiment}) and repeating the spectral measurements the sample transmittance can be computed as 
\begin{equation} 
T=\frac{R_{S}}{R_{BG}}
\label{eq:transmittance}
\end{equation}
where 
\begin{equation}
    R_{S}=\frac{S^{S}_{S}}{S^{S}_{{N}}}
\end{equation}
 and $S^{S}_{{S}}$ and $S^{S}_{{N}}$ are the spectra recorded with detector $D_S$ and $D_N$, respectively. 
We note that in the system configuration shown in Fig.~\ref{fig:experiment} only the magnitude information, i.~e.~transmittance or absorbance, can be measured. However, arranging the system geometry as in the case for Dispersive FT spectroscopy \cite{parker_dispersive_1990}, where a sample is place in one of the arms of the interferometer, the phase information can be obtained as well. Since in our case we utilize a normalisation detector for common intensity noise suppression of the QCL, a scheme as proposed by Schiller \cite{schiller_spectrometry_2002} for dual-comb spectroscopy, where the second comb corresponds in our case to the Doppler shifted beam path, can be exploited as well. 

\subsection{Amplitude noise of the system}
\begin{figure}[h]
    \centering
    \includegraphics[width=1.0\textwidth]{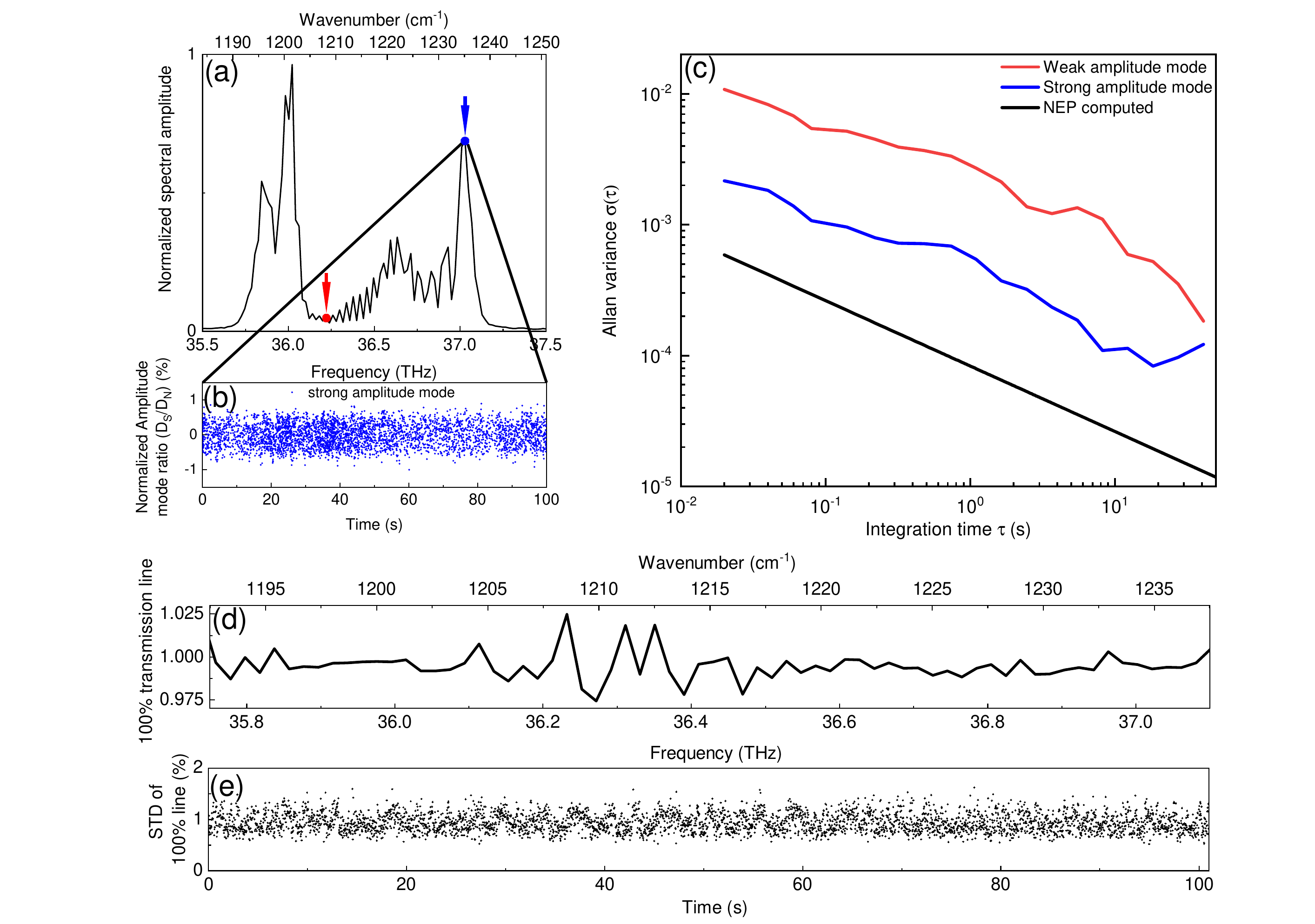}
    \caption{\textbf{Amplitude noise investigation. (a) QCL spectrum on which Allan deviation is performed. (b) Temporal evolution of normalized amplitude mode ratio of the strongest mode, which is marked with blue arrow in (a). It shows the system stability over the time period of 100s with amplitude fluctuations below 1\%. (c) Computed Allan variance of high (blue) and low (red) spectral amplitudes of the spectrum in (a) which is indicated with arrows in the same color code. The black curve shows the computed Allan variance taking the background (Noise equivalent power (NEP) of a detector) into account for a power of strong mode. (d) Background, or so called 100\% line transmission, of a system for a single interferogram and its temporal stability expressed as a standard deviation (STD) in \%  over the entire optical bandwidth  over 100s}}
    \label{allan_plots_1}
\end{figure}

Before demonstrating high-resolution spectroscopy we analyse the system sensitivity and discuss possible noise sources and their origin. The SNR of the system is given by the amplitude noise of the heterodyne beat note signal in the RF domain. For this purpose we analyse the amplitude noise fluctuations of the spectrum shown in Fig.~\ref{allan_plots_1} (a). Fig.~\ref{allan_plots_1} (b) shows a time trace of the normalized amplitude ratio of the strong mode, with an amplitude fluctuations  $<1\%$, showing excellent stability over several minutes. To quantify different noise sources we compute the Allan deviation $\sigma(\tau)$ of comb teeth with high and low amplitudes, indicated in Fig.~\ref{allan_plots_1} (a) with red and blue arrows. Fig.~\ref{allan_plots_1} (b) shows the Allan deviations as functions of the integration time $\tau$. Additionally, we show the computed noise floor originating from the detector noise-equivalent power (NEP) computed from the spectrum for a power of a strong mode. Both the high and low amplitude peaks show that the noise of the system decreases with the square root of the integration time $\tau$ up to 40 s and the system SNR is close to the detector NEP which deviates by a factor of 2.8 from the NEP limit. This deviation might originated from additional thermal and electrical noise sources which are not take into account by computing NEP. Thus, with the employed detectors and the laser, the SNR can be improved by longer integration times or by more powerful laser, since the detector saturation power is around 8 mW, whereby the average power on the detector is 0.5 mW. With higher power close to the saturation power we would expect improvement of the SNR by factor 10. At longer integration times ($\tau >$10 s) we can reach SNR of 10$^4$ and 10$^3$ for strong and weak modes, respectively. We would like to point out that the 1/f noise in the system is minimized, due to fast acquisition, since we observe a slope of $-1/2$ for entire Allan plot which is characteristic of NEP-noise-limited systems. In order to perform long-term measurements, we have to quantify the background stability. For this reason we record the system background on minutes time scales. Fig.~\ref{allan_plots_1} (d) shows a normalized system background ($R_{BG}$) spectrum, so called 100$\%$ transmission line, for a single interferogram over the entire comb bandwidth. Since the comb spectrum consist out of two lobs with high power on the sides and with low power in the center of the spectrum, the computed 100\% line also reflects this spectral power distribution with 2\% deviation in the weak spectral region. For demonstration of the system and hence background stability we show the computed standard deviation (STD) of the 100\% line over the entire spectrum in Fig.~\ref{allan_plots_1} (e). As we can see from the Fig.~\ref{allan_plots_1} (e) the system background remains stable over 100 seconds with deviation below 1.5\%. This ensures for performing long-term measurements which are characteristic for high resolution spectroscopy via interleaving technique \cite{gianella_high-resolution_2020}.             

\subsection{Interleaving spectroscopy \label{Interleaving}}\label{Interleaving}
\subsubsection*{Interleaving on silicon etalon}
In order to demonstrate the system capabilities in high-resolution spectroscopy, we use spectral interleaving similar to what described in Ref. \cite{gianella_high-resolution_2020}. In the case of interleaving, the resulted point spacing resolution is smaller than the free spectral range of the laser  given by the QCL cavity length, which in our case corresponds to a resolution of $\approx$ 19.7 GHz. In contrast to dual-comb spectroscopy, where two synchronised current ramps with different slopes are required for the laser tuning \cite{gianella_high-resolution_2020}, we apply a single triangular current modulation ramp (Fig.~\ref{fig:experiment}) to the QCL current driver (Wavelength electronics QCL1000) which is controlled by means of a wave function generator. The interferogram acquisition trigger (the Hall sensor of the motor) does not need to be synchronized with the wave function generator, since the extracted frequencies from individual spectra can be sorted in post-processing. The triangular current ramp of the function generator is set so that the overall QCL spectrum can be tuned by one $f_{rep}$. In this way, we can access frequencies between modes and even provide a gap-less spectral coverage over the entire optical bandwidth of the QCL comb. Firstly, we demonstrate the high-resolution capability of the system on $\approx$500 $\mu$m thick silicon etalon. For this measurement, we electrically tune the QCL spectrum over one $f_{rep}$ in about 7 seconds. During this time, the measurements are acquired continuously at a rate of 50 Hz. Due to the slow current ramp rate and fast interferogram acquisition rate (1 ms) with 20 ms periodicity (acquisition on the same single retro-reflector), the information within the recorded single-burst interferogram is not influenced by the slow current changes induced by the current ramp. As explained earlier, we record a background and a sample measurement, and extract information from a single-burst interferogram via the sub-nominal resolution technique \cite{rutkowski_optical_2018}. The corresponding interleaving data is shown in Fig.~\ref{interleaving_new} (a), where 20 consecutive interferograms are coherently co-added before a spectrum is computed. The resulting spectra are then frequency binned to 5 GHz resolution. As it can be seen in Fig.~\ref{interleaving_new} (a) and the zoomed region (Fig.~\ref{interleaving_new} (b)) the recorded interleaved spectra fit well the computed etalon spectrum. Measurements deviation from the model, especially at frequencies with high transmission can be attributed to losses in the silicon etalon which were not taken into account in the model. Also small angle deviation from perpendicular incidence on the etalon will result in deviation from the model. This measurement shows that in a relatively short time period of 7 s, we can clearly resolve etalon features with resolution $<$0.15cm $^{-1}$. This addresses the question whether sub-GHz spectroscopy is possible with our spectrometer.                 
\subsubsection*{Interleaving on methane} \label{Methane}
\begin{figure}[h]
    \centering
    \includegraphics[width=5.2in]{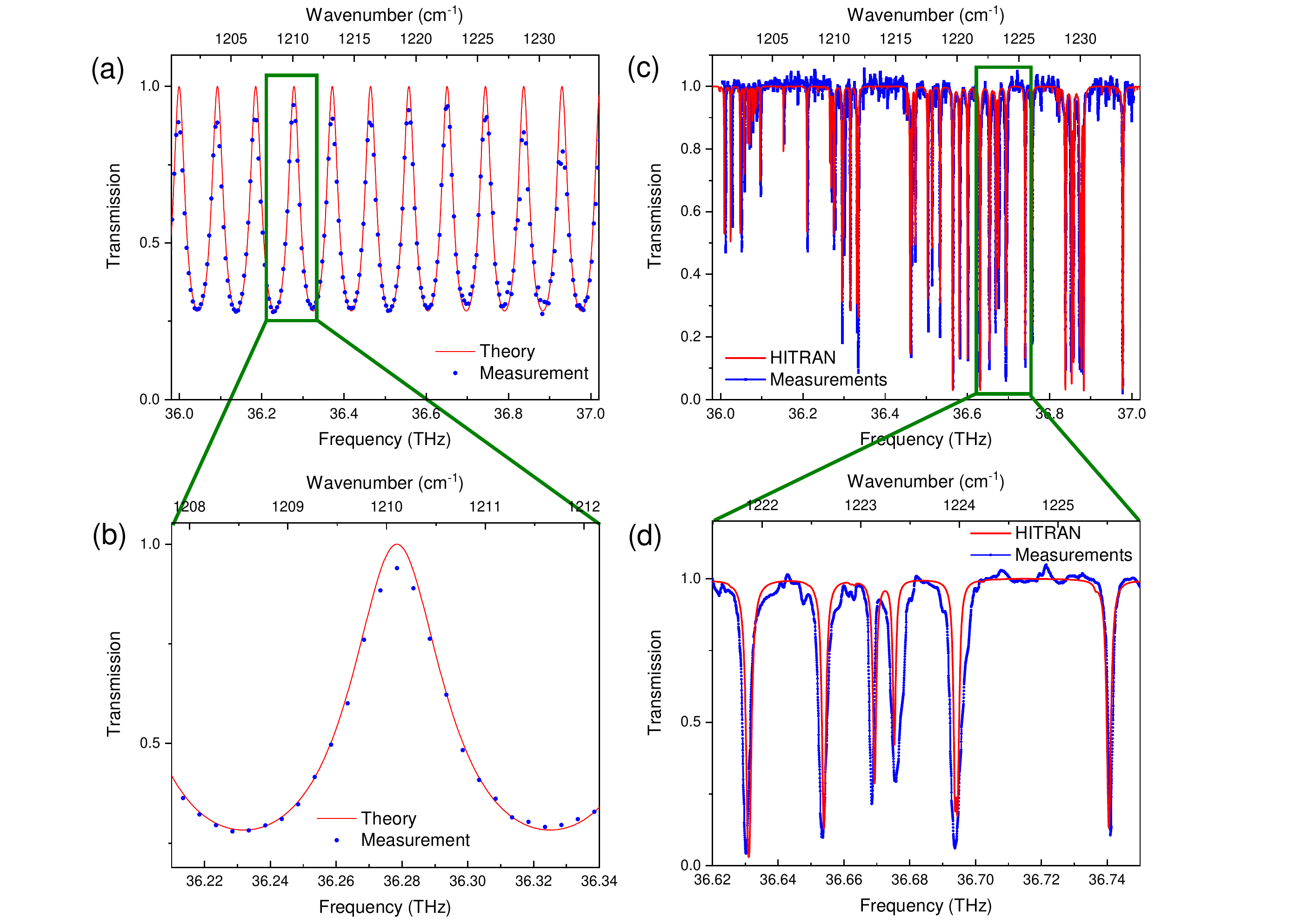}
    \caption{\textbf{High resolution spectroscopy. (a)/(b) Interleaving spectrum (blue) of $\approx$ 500$\mu$m thick silicon etalon with a binned frequency resolution to 5 GHz and the computed etalon transmission (red) and the zoom in recorded in 7 seconds. (c) Doppler broaden methane (CH$_4$) spectrum recorded via interleaving (blue) at a pressure of 200 mbar in 25 seconds with frequency binned resolution to 250 MHz and the HITRAN data base reference (red). (d) Zoom in of (c) with sub-GHz resolved methane absorption lines.}}
    \label{interleaving_new}
\end{figure}

A more challenging and practically relevant scenario is high-resolution spectroscopy on low-pressure Doppler-broadened methane lines. For this experiment, the background measurements were carried out with a 10 cm-long evacuated gas cell inserted into the sample beam path and the sample measurements were performed on the filled gas cell with CH$_4$ at a pressure of 200 hPa. Contrary to the low-finesse etalon in the previous section, the low pressure methane lines have sub-GHz linewidths. The high-resolution measurements were performed in the same way as for the etalon, but with the QCL current ramp period set to 25 seconds. In this way, thousands of interferograms are recorded, whereby each interferogram is recorded on ms time scale, and 10 consecutive ones are co-added before computing each spectra. With such a slow current ramp period we can ensure a high frequency resolution and resolve low pressure Doppler broadened methane lines. The resulting interleaved spectrum of methane is shown in Fig.~\ref{interleaving_new} (c), where the resolution has been reduced by frequency binning to 250 MHz. There are two reasons for frequency binning to 250 MHz. First, it results in improved SNR and secondly the interleaved spectrum is equidistant. The zoom into the spectrum of Fig.~\ref{interleaving_new} (c) is shown in Fig.~\ref{interleaving_new} (d) and shows that the system is capable of resolving Doppler broadened molecular absorption lines with sub-GHz linewidths. However, the resolved linewidths of some methane lines are broadened by up to 5\% compared to the bare HITRAN 2020 data base spectrum \cite{gordon_hitran2020_2022}. For the computed HITRAN data spectrum no instrumental linewidth has been taken into account, because we do not expect any line broadening from our rotational FT spectrometer in combination with the QCL frequency comb. Additionally, we observe that the 100\% transmission line is not entirely flat. This measurement deviations can be explained with contributions from two different origins. Firstly, the QCL comb short-term stability can be affected by back-reflections in the system. Secondly, high-frequency electrical instabilities of the laser can occur while interleaving, since it is challenging to identify a  current range where the QCL tunes continuously over one $f_\text{rep}$ in the comb regime. This will result in slightly different QCL states for background and sample measurement. We note that this noise source could be further reduced by using low-noise current sources and a more stable comb source. However, Fig.~\ref{interleaving_new} (d) shows clearly that all sub-GHz CH$_4$ absorption lines can be well resolved. Providing sub-GHz spectral resolution shows a clear advantage of our approach, since such a high resolution is challenging to achieve with the sate-of-the-art FT spectrometers which are typically on the order of 0.2 cm$^{-1}$ (6 GHz) or 0.06 cm$^{-1}$ (1.8 GHz) for high resolution spectroscopy. 
\section{Time-resolved measurements on methane}
\begin{figure}[h]
    \centering
    \includegraphics[width=4.7in]{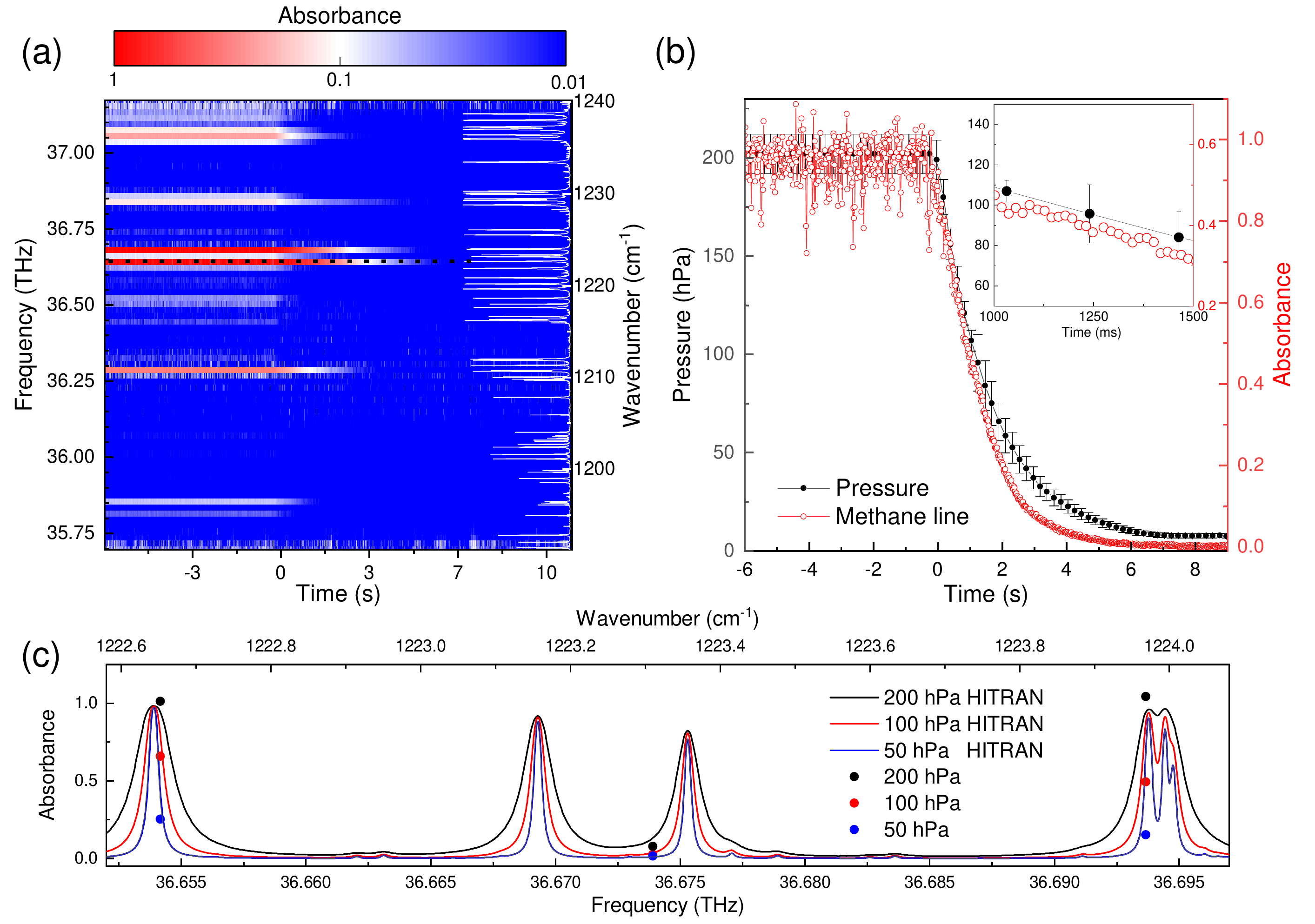}
    \caption{\textbf{Time-resolved spectroscopy on methane. (a) Absorbance is computed from time resolved-transmission measurements on the filled gas cell to a pressure of 200 hPa which is evacuated via a pump. The computed methane absorbance for a pressure of 200 hPa is shown on the right side with white lines. Temporal evolution is monitored by initializing the evacuation of the gas cell a $t=0$ s. The acquisition is performed with periodicity of 10 ms in the comb regime at a constant current mode. Simultaneously several low pressure methane lines are observed for $t<0$ s. (b) Visualization of the temporal evolution of the strongest absorption line (marked with a dotted black line) in Fig. (a) and the corresponding pressure of the gas cell. A zoom in is shown in the top right corner with 500 ms time duration starting 1s after opening the valve. (c) Computed HITRAN absorbance of methane at a pressure of 200 hPa, 100 hPa and 50 hPa and the measured absorbance close to the strongest absorbance peak at these pressure and hence corresponding laboratory time.}}
    \label{absorbance}
\end{figure}

Replacing the translational delay stage used in state-of-the-art FTIR spectrometers by our rotational delay line enables much higher acquisition speeds for a given spectral point spacing. In order to demonstrate this capability of our system, we perform time-resolved spectroscopy on methane. We monitor multiple methane absorption lines simultaneously while reducing the pressure of the methane in the cell over a few ($<$6) seconds. Initially, a gas cell is filled with methane at a low pressure of 200 hPa via a gas inlet attached to a pump via mechanical valve, while the pressure is monitored with a vacuum gauge (Pfeiffer vacuum gauge controller TPG 361) in 100 ms time steps via a second connection. At $t = 0$ the valve is opened while simultaneously the pressure and the absorbance are monitored. The absorbance spectra are acquired in $\sim$10 ms intervals by acquiring single interferograms in 625 $\mu$s with the QCL operated at a constant current in the comb regime. From the recorded transmittance ($I/I_0$,) and the background measurement (empty cell), we compute the absorbance \begin{equation}
A=-\log_{10}(I/I_0),
\end{equation} which is shown in Fig.~\ref{absorbance} (a). For $t<0$ s, where the valve is closed, we observe 10 methane absorption lines within the optical bandwidth of the comb. By opening the valve ($t=0$ s) we observe a rapid absorbance decrease within 2 seconds. The absorbance is given by the Beer–Lambert law as
\begin{equation}
A=\epsilon_{\lambda}\cdot \tilde{c}\cdot l,
\end{equation} 
where $\epsilon_{\lambda}$ is the molar attenuation coefficient at wavelength $\lambda$, $\tilde{c}$ the amount concentration, and $l$ the length of the gas cell. Both $\epsilon_{\lambda}$ and $\tilde{c}$ are functions of the gas pressure $p$ and temperature. Since the linewidth of methane absorption lines (sub GHz) at this low pressure is much smaller than the spectral point spacing of $\sim 19.7$ GHz of the particular QCL comb used in this experiment, the absorbance is measured off the center of the absorption lines. This results in a non-trivial time-dependence of the absorbance as the molar attenuation coefficient is both wavelength and pressure dependent; namely, a linewidth narrowing of the Doppler broaden methane absorption lines and hence in the time-dependent change of $\epsilon_{\lambda}$, as well as a decrease in the amount concentration $c(p)$. We demonstrate this effect by showing a cut in Fig.~\ref{absorbance} (a) (cut along the marked dotted line) along the temporal-evolution of the strongest absorbance line in Fig.~\ref{absorbance} (b), and compare it to the monitored pressure inside the gas cell. The absorbance and the pressure decrease at the same rate down to a pressure of about 100 hPa. Moreover, for $t>6$ s, which corresponds to a pressure of 7 hPa, the absorbance is zero, which can be explained by the fact that the QCL mode is either off from the resonance of the methane absorbance line, or it is located in the absorption tail, thus resulting in a weak absorption. We show the pressure related linewidth narrowing of methane in Fig.~\ref{absorbance} (c) where we compute the methane absorbance exemplary for three different pressures and overlap the measured absorbance points which corresponds to measurements times when these pressure are reached. We observe a good agreement of spectral and temporal resolved methane absorbance lines. Please note, that each absorbance line can be seen as independent barometer, hence by calibrating each line one can increase the accuracy of the measurement. Our spectrometer demonstrates high temporal and spectral resolution which is predominant over the state-of-the-art FT spectrometers, which are not able to provide such simultaneous temporal and spectral resolution.

\section{Discussion}
In this work we have demonstrated a proof-of-principle of the rotational FT spectrometer which decouples the spectral resolution from the temporal one. Moreover, we have demonstrated millisecond dual-comb spectroscopy using a single QCL frequency comb, enabled by the fast acquisition speed of the system. A further benefit of a fast acquisition is the minimized 1/f noise within the signal bandwidth, which is inevitable in the state-of-the-art FTIRs. This is due to the fact, that the RIN of a free running laser is reduced at least by two to three orders of magnitude \cite{cappelli_intrinsic_2015}. Additionally, we have demonstrated high resolution, broadband spectroscopy capabilities on a low-finesse etalon and low-pressure, Doppler-broadened methane lines via the interleaving technique. Furthermore, since the geometry of the rotational delay line allows incorporation of multiple sources, either via time-multiplexing scheme or under simultaneous operation, a broader optical bandwidth can be achieved straight-forwardly. We would like to point out that our developed system may be exploited for the recently proposed frequency comb ptychoscopy scheme \cite{benirschke_frequency_2021}, where an external signal is analysed with a frequency comb and a Doppler-shifted version of it. Since the frequency shift of the Doppler shifted comb can be adjusted to acoustical frequencies (sub MHz), the system can also be employed for photo-acoustic dual-comb spectroscopy \cite{wildi_photo-acoustic_2020}. 
Moreover, the geometrical structure of the developed rotational delay line is attractive for the implementation of multidimensional FT spectroscopy \cite{lomsadze_frequency_2017} with a single frequency comb source. The compact system demonstrated in this work is of high interest for performing high spectral resolution and time-resolved spectroscopy, especially when combining with ultra broad band white, NIR and mid-IR supercontinuum laser sources \cite{hoghooghi_1-ghz_2022,alfano_supercontinuum_2016,chang_ultra-broadband_2021,diouf_ultra-broadband_2017}. Moreover, in the case of light sources with stable temporal intensity, the complexity of the system can be  reduces by eliminating the normalization detector and reducing the acquisition system to two channels. The nominal spectral resolution can be significantly improved by a factor 2 to 4 by increasing the size of the RDL or by increasing the number of reflections ($N$ in Eqn.~\eqref{eqn:path}). An increased temporal resolution is also readily achievable by replacing the existing motor with a magnetically levitated one. This would increase the single-spectrum acquisition speed up to 30 kHz (33$\mu$s), which is sufficient for the majority of spectroscopy applications, such as reaction monitoring \cite{lehman_optical_2021}, leakage monitoring \cite{alden_single-blind_2019} and in-line process analytical technology (PAT) \cite{jin_near_2013}.

\newpage

\backmatter

\bmhead{Supplementary information}
Supplementary information is provided additionally.  
\bmhead{Acknowledgments}
We acknowledge the NCCR QSIT, the ERC Grant CHIC (No. 724344), the quantum Flagship project QOMBS and the EU project FLASH for financial support. G.S. would like to thank J.Hillbrand for help and A. Mayer for discussions.

\bmhead{Competing Interests} The authors declare no competing interests.

\bmhead{Author contributions}G.S. conceived the idea for Doppler-based spectroscopy. S.M. invented, designed, and constructed the rotational FT spectrometer, and carried out the measurements. S.M and M.F performed the data analysis. M.F. implemented a real-time data acquisition on a graphical processor unit (GPU). Early prototypes for rotational delay lines were investigated by S.M., M.S. and A.F. Initial laser characterisation were performed by M.Bertrand. Laser was processed by P.J. and grown by M.Beck. The manuscript was written by S.M and M.F. All authors provided critical feedback and contributed to the final shape of the manuscript. The work was supervised by J.F.and G.S.

\newpage

\bibliography{sn-bibliography}

\end{document}